\documentclass[11pt]{article}

\usepackage[style=ieee,backend=biber]{biblatex}
\usepackage{graphicx}
\usepackage{tabularx}
\usepackage{booktabs}
\usepackage[T1]{fontenc}
\usepackage{lmodern}
\usepackage[flushleft]{threeparttable}
\usepackage[hidelinks]{hyperref}
\hypersetup{hidelinks}
\usepackage{multirow}
\usepackage{longtable}
\usepackage[margin=3cm]{geometry}
\usepackage{amsmath}
\usepackage{amssymb}
\usepackage{authblk}
\usepackage{caption}
\usepackage{flafter}
\title{Over-the-air White-box Attack on the Wav2Vec Speech Recognition Neural Network}
\author[1]{Alexey Protopopov}
\affil[1]{Joint Stock Research and Production Company Kryptonite \authorcr
E-mail: a.protopopov@kryptonite.ru}
\date{}
\begin{document}
    \captionsetup[figure]{labelformat={default},labelsep=period,name={Figure}}
    \captionsetup[table]{labelformat={default},labelsep=period,name={Table}}

    \maketitle

    \begin{abstract}
	Automatic speech recognition systems based on neural networks are vulnerable to adversarial attacks that alter transcriptions in a malicious way. Recent works in this field have focused on making attacks work in over-the-air scenarios, however such attacks are typically detectable by human hearing, limiting their potential applications. In the present work we explore different approaches of making over-the-air attacks less detectable, as well as the impact these approaches have on the attacks' effectiveness.
    \end{abstract}


    \section{Introduction}\label{introduction}

In the past few years there has been an effort to study the vulnerability of neural networks (NNs) to adversarial attacks: specially constructed inputs designed to make the NN produce malicious results, e.g. misclassify an object. Since the overwhelming majority of modern automatic speech recognition (ASR) systems is based on NNs, this vulnerability extends to them. Many of the existing attacks are designed to work over-the-wire \autocite{Carlini,Schoenherr,Qin}, i.e. they exist as an audio file that is intended to be uploaded into an ASR system. Such an approach has limited practical value, so a significant effort is being directed into producing an attack that would work over-the-air, meaning that it would remain effective after being played over a speaker and picked up by a microphone. While there are successful examples of such attacks \autocite{Ettenhofer,Fang}, they cannot be considered truly imperceptible. In the present work, we attempted to design an algorithm that could generate a reliable and imperceptible over-the-air attack. We have found that implementing measures to reduce the attacks’ perceptibility tends to have an unexpected but welcome side effect of improving their robustness, but at the same time drastically increases the amount of computations needed to generate the attack. 

\section{Methods}\label{methods}

We used the Wav2Vec2 model \autocite{Baevski} as our target ASR system since it is open-source and reasonably modern. Audio files were played back using a JBL Go 3 portable speaker and received using the built-in microphone of a Lenovo ThinkPad laptop.
On the most basic level, the operation of the ASR system can be described by the following formula:

\begin{equation}
    T(x)=t
    \label{eq:1}
\end{equation}

where \emph{x} is the original audio, \emph{T} is the ASR system presented as a transform and \emph{t} is the resulting transcription. Likewise, an attack can be written as follows:

\begin{equation}
	\begin{aligned}
	    &T(x')=t' \\
	    &x'=x+\Delta
	    \label{eq:2}
    \end{aligned}
\end{equation}

where \emph{$\Delta$} is the modification applied to the original audio, and \emph{t'} is the transcription that one would want to get from the modified audio \emph{x'}, where  $\emph{t'}\neq\emph{t}$. In this notation, generating the attack would imply finding the appropriate \emph{$\Delta$}. Carlini et al. in \autocite{Carlini} have proposed a method involving backpropagation from a loss function to optimize \emph{$\Delta$} over a number of iterations. However, in order to make the attack work over-the-air we needed to introduce certain modifications to the original method. Specifically, several data augmentation steps were introduced:

\begin{enumerate}
    \item Psychoacoustic masking (PAM);
    \item Speaker and microphone frequency response (FR) simulation;
    \item Room simulation;
    \item Random audio time shifts.
\end{enumerate}

\begin{figure}[htbp]
    \centering
    \includegraphics[scale=0.6]{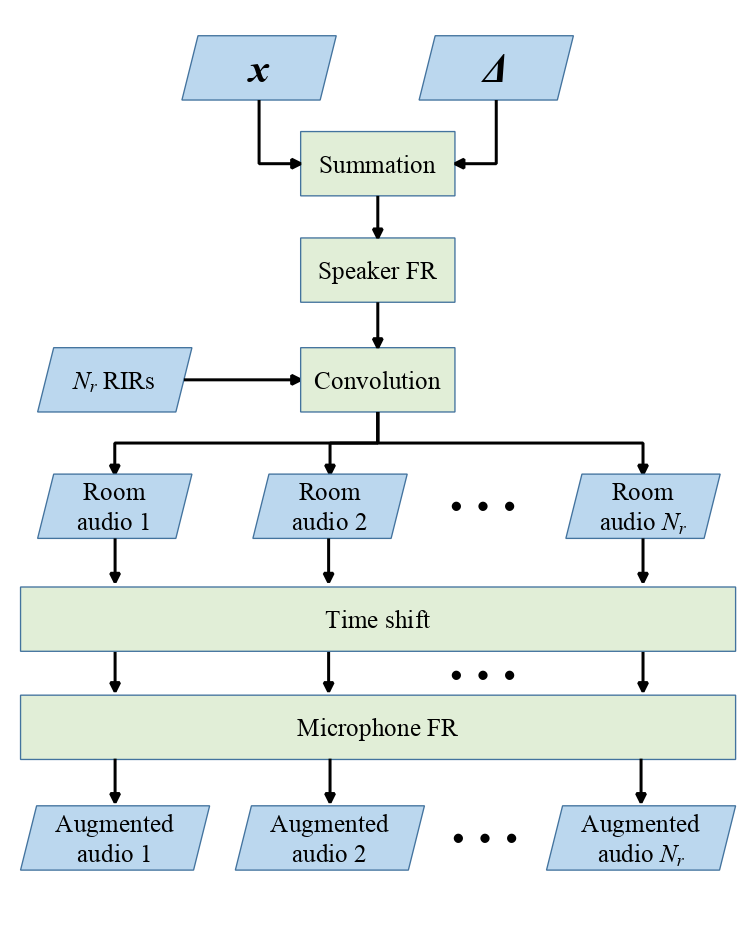}
    \caption{Data augmentation.}
    \label{fig:1}
\end{figure}

These augmentation steps were applied to the audio at every iteration. The first item on this list, PAM, is of particular interest and is based on the approach suggested in \autocite{Schoenherr}. It exploits a trait of human hearing, where a tone of frequency \emph{$\nu\textsubscript{0}$} can mask weaker tones of frequencies 
$\nu\textsubscript{0}\pm\Delta\nu$
provided \emph{$\Delta\nu$} is small enough. This means that given a spectrogram \emph{X} of waveform \emph{x}, we can use it to calculate a matrix \emph{M\textsubscript{PAM}} with the same dimensions as \emph{X}, which would contain hearing thresholds for each frequency in each spectrogram time window. Theoretically, if \emph{$\Delta$} remains below these thresholds, it should be imperceptible to human hearing. Thus, we add this as a condition to our loss function:

\begin{equation}
    L(\Delta)=L\textsubscript{CTC}(x+\Delta)+\lambda \cdot f\textsubscript{PAM}(\Delta)
    \label{eq:3}
\end{equation}

where \emph{L\textsubscript{CTC}} is the connectionist temporal classification (CTC) loss and \emph{f\textsubscript{PAM}} is the function of the difference between \emph{$\Delta$} and \emph{M\textsubscript{PAM}}, which is calculated as follows:
\begin{equation}
    f\textsubscript{PAM}(\Delta)=||\textrm{ReLU}(F(\Delta)-M\textsubscript{PAM})||\textsubscript{2}
    \label{eq:4}
\end{equation}

where \emph{F($\Delta$)} is a fast Fourier transform applied to \emph{$\Delta$}. In essence, we subtract the threshold matrix from the spectrogram of \emph{$\Delta$}, set negative values to zero to prevent the optimizer from reducing the amplitude of \emph{$\Delta$} below what is needed, and then take an \emph{L\textsubscript{2}} norm of the resulting matrix. This results in a scalar which is introduced into the loss function with a scaling coefficient \emph{$\lambda$}, which is used to control the degree of masking we want to achieve. Overall, the data augmentation can be summarized using the schematic on Figure~\ref{fig:1}.

\subsection{Frequency response simulation}\label{freq_resp_sim}

All speakers and microphones have specific FRs, thus it makes sense to simulate them in our attack generation process. The exact FRs vary from one device to another, so we decided to generalize it by using a bandwidth filter with a pass band of [50;~8000]~Hz. Conveniently, the FRs of most commercial speakers and microphones are not too dissimilar, so this filter can be used to simulate both, and is thus applied twice: before and after room simulation.

\subsection{Room simulation}\label{room_sim}

\begin{figure}[htbp]
    \centering
    \includegraphics[scale=0.8]{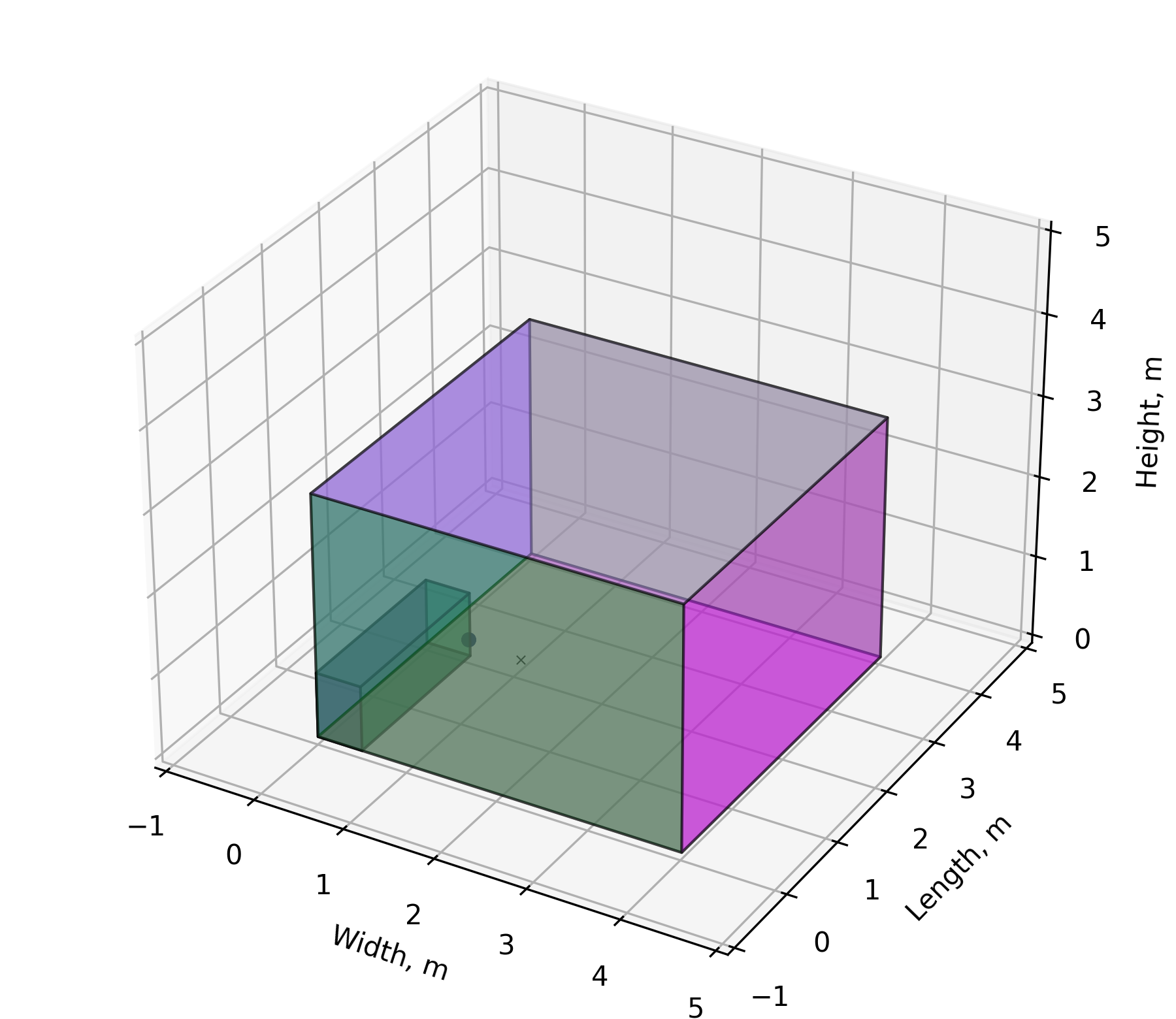}
    \caption{Example room layout.}
    \label{fig:2}
\end{figure}

Because we expect the attack to work in a real environment, we need to account for the distortions caused by the configuration of the room the attack takes place in. We used the Pyroomacoustics software package, which simulates sound reflection off solid objects by using virtual sound sources. The result of this simulation is a room impulse response (RIR), which is convolved with the result of the speaker frequency response simulation. Naturally, it is impossible to predict the exact layout of the room where the attack will be used, so our strategy was to make an attack that would be equally effective in a sufficiently large variety of different environments. For this purpose we have created a room generation algorithm, which takes a room template and changes some of its parameters to produce an arbitrary number of room variants. The variable parameters included the positions of the walls, microphone and speaker. An example of one such room can be seen in Figure~\ref{fig:2}.

\begin{figure}[htbp]
    \centering
    \includegraphics[scale=0.6]{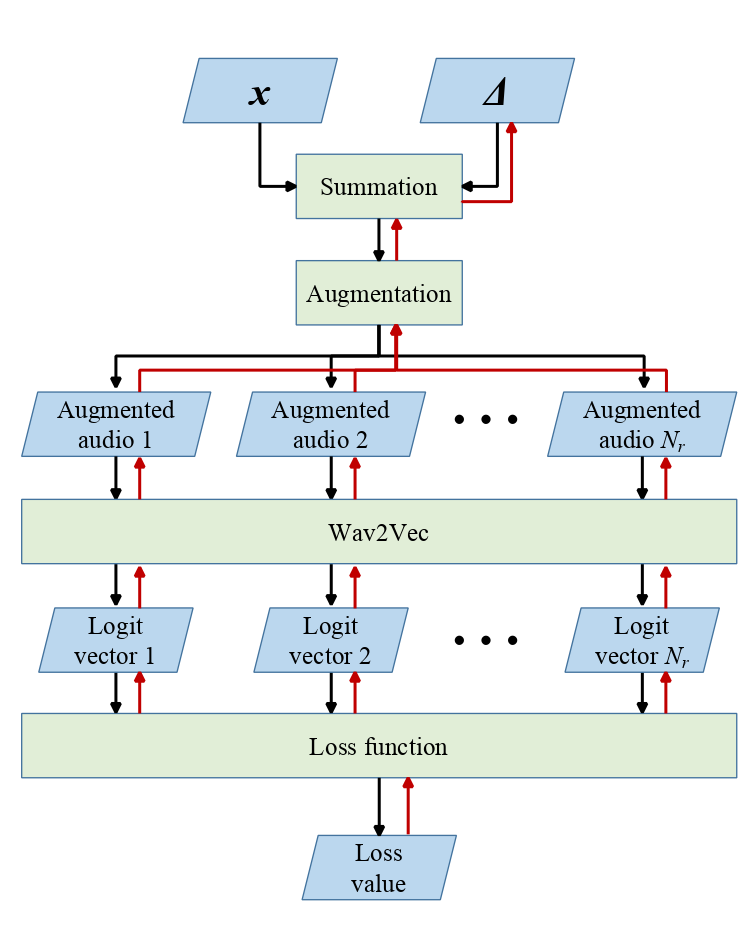}
    \caption{Attack generation iteration. Gradient backpropagation is indicated by red arrows.}
    \label{fig:3}
\end{figure}

All rooms had a parallelepiped shape with a smaller parallelepiped inside to simulate a sofa. Appropriate materials were selected for the surfaces of the simulated rooms, in particular, one of the walls was assigned a cloth material to simulate a curtain. Over 700~rooms were generated this way, and all of them were used simultaneously to generate the attack. This means that each of the \emph{N\textsubscript{r}} simulated RIRs was convolved with the result of the speaker frequency response simulation to produce \emph{N\textsubscript{r}} audio waveforms, each of which was then subjected to all subsequent data augmentation steps, passed through the ASR system and introduced into the loss function to produce \emph{N\textsubscript{r}} scalars which are summed up to produce a single value which serves as a starting point for gradient backpropagation, as shown on Figure~\ref{fig:3}.

Thus, at every iteration the gradients propagate through every room involved in the generation process and the effect of samples corresponding to different rooms are added together.

\subsection{Random audio time shifts}\label{time_shifts}

\begin{figure}[htbp]
    \centering
    \includegraphics[scale=0.4]{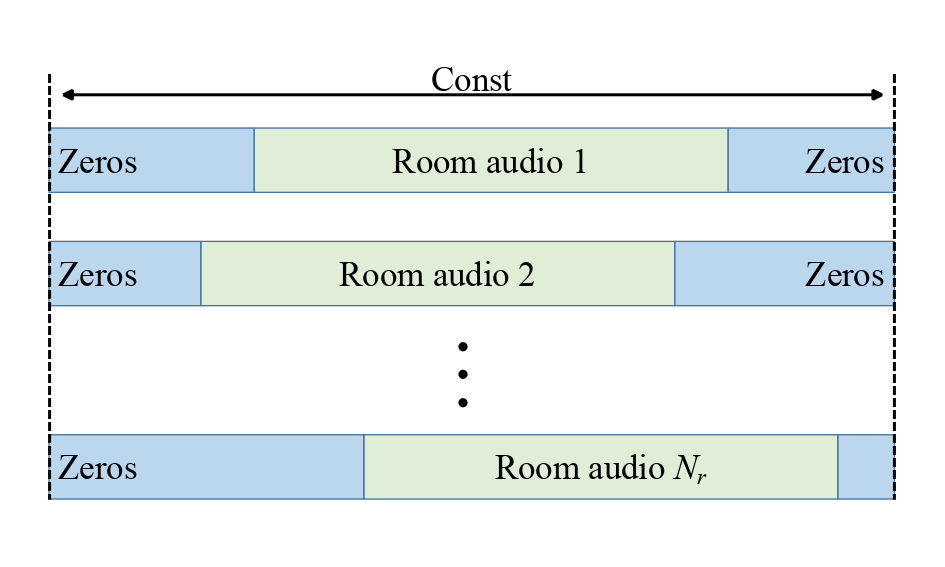}
    \caption{Audio fragments padded with zeros.}
    \label{fig:4}
\end{figure}

Wav2Vec breaks down the input audio into a series of windows to assign phoneme labels to each of them, which means that we cannot guarantee that the alignment of these windows will be the same during generation and during application, since the attack is intended to work over the air. Thus, it has to be made insensitive to time shifts, which is done by padding both ends of the modified audio \emph{x'} with sequences of zeros of random lengths, as shown on Figure~\ref{fig:4}. The padding lengths are randomized at every iteration.

\subsection{Attack generation}\label{attack_generation}

We generated the attacks by iterating as shown on Figure~\ref{fig:3} while the transcription of the attack \emph{T(x')} was compared against the target transcription \emph{t'} at every 10 iterations for all \emph{N\textsubscript{r}} simulated rooms. We would start at \emph{$\lambda$}=0 and increase it by a fixed amount every time there is a total transcription match (i.e. in all \emph{N\textsubscript{r}} cases). The generation would stop once \emph{$\lambda$}=1 and there is a total transcription match.

\section{Results and discussion}\label{results}

The attacks generated by the algorithm were tested by playing them over a JBL Go 3 portable speaker, recording the sound in WAV format using the built-in microphone of a Lenovo ThinkPad laptop and then transcribing them using Wav2Vec ASR system. The following parameters were used to gauge the effectiveness of the attack: success rate, phoneme error rate (PER) and word error rate (WER). An attack was considered successful if its transcription perfectly matched the desired transcription. Each experiment consisted of us playing back and recording the same attack audio 10~times and then averaging the measured parameters. The results of these experiments are presented in Table~\ref{tab:1}. In addition to evaluating the performance of the attack at various \emph{$\lambda$}, we also tested the performance with some data augmentation steps omitted.

\begin{table}
    \caption{Experimental results. FR — frequency response, RS — room simulation, TS — time shift.}
    \label{tab:1}
    \centering
    \begin{tabular}{||c|c|c|c|c|c|c|c|c||}
        \hline
        Experiment & FR & RS & TS & \emph{$\lambda$} & Success rate & PER & WER & Generation time, hrs \\
        \hline
        1  & + & + & + & 0.00 & 0\% & 0.13 & 0.44 & 1.0 \\
        \hline
        2  & + & + & + & 0.15 & 40\% & 0.03 & 0.09 & 11.5 \\
        \hline
        3  & --- & + & + & 0.00 & 0\% & 0.20 & 0.45 & 1.4 \\
        \hline
        4  & --- & + & + & 0.15 & 20\% & 0.06 & 0.20 & 8.2 \\
        \hline
        5  & + & --- & + & 0.00 & 0\% & 0.94 & 0.86 & 0.3 \\
        \hline
        6  & + & --- & + & 0.15 & 0\% & 0.91 & 0.87 & 1.1 \\
        \hline
        7  & + & + & --- & 0.00 & 0\% & 0.55 & 0.83 & 0.5 \\
        \hline
        8  & + & + & --- & 0.15 & 10\% & 0.10 & 0.24 & 5.0 \\
        \hline
    \end{tabular}
\end{table}

\begin{figure}[htbp]
    \centering
    \includegraphics[scale=0.8]{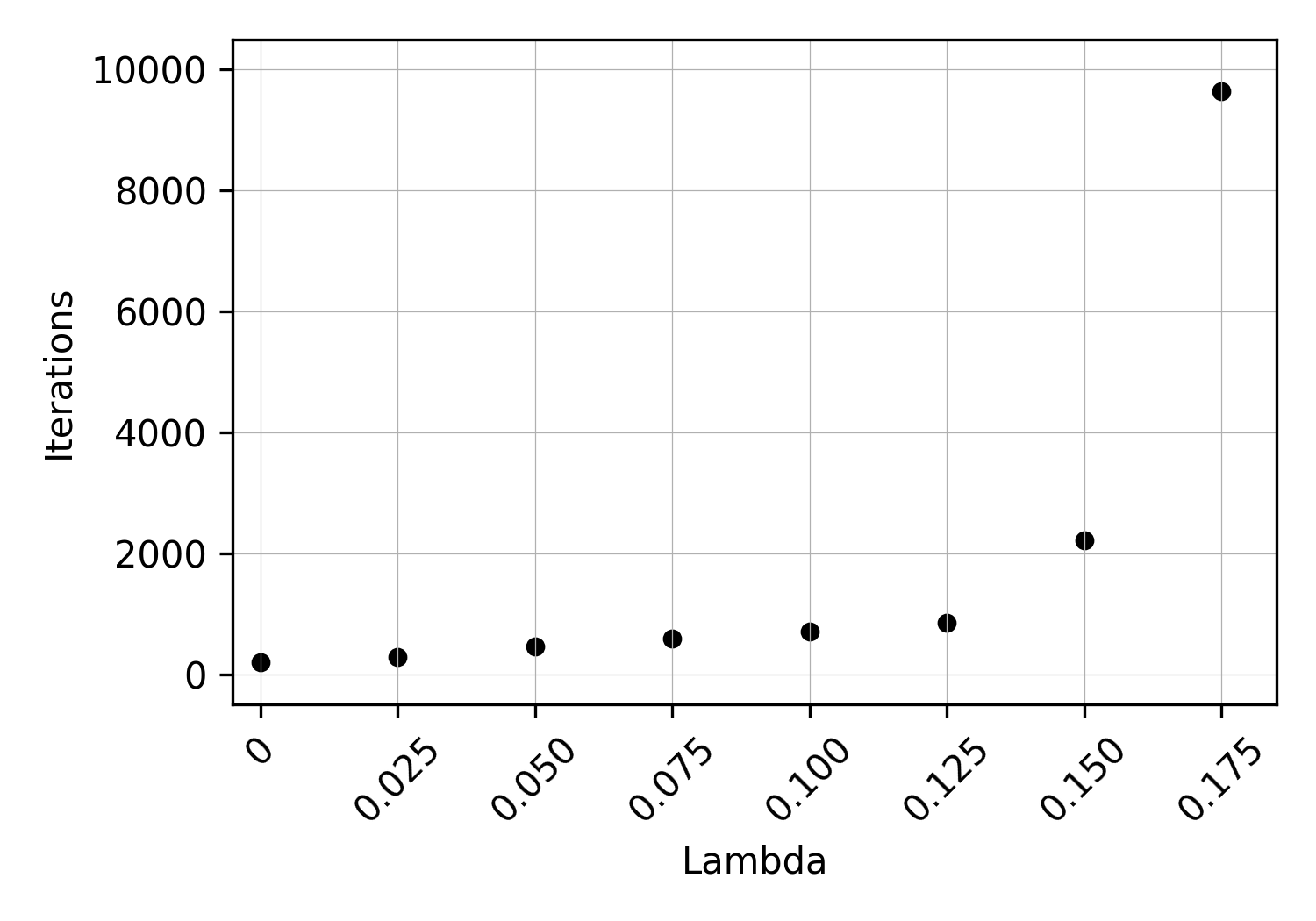}
    \caption{Numbers of iterations required to reach various values of \emph{$\lambda$}.}
    \label{fig:5}
\end{figure}

Since PAM only serves to reduce the attack’s overtness to the human ear by reducing it’s overall magnitude, it is natural to expect the success rate to decline as \emph{$\lambda$} increases. However, increasing \emph{$\lambda$} had an unexpected side effect of increasing the attack’s effectiveness as well. Unfortunately, generating an attack with a higher \emph{$\lambda$} requires a significantly larger number of iterations, as can be seen in Figure~\ref{fig:5}. With the entire set of augmentation steps enabled, performing 10,000 iterations took just over 2 days on the GPU available to us. Given the trend shown in Figure~\ref{fig:5}, we decided not to generate the attacks any further as we were no longer confident that the generation process would reach any meaningful result in the foreseeable future.

\section{Conclusion}\label{conclusion}

The proposed algorithm allows generation of adversarial attacks against Wav2Vec ASR system, which have been demonstrated to be successful in over the air experiments. Our efforts to conceal these attacks have paid off, to an extent, as increasing \emph{$\lambda$} did indeed make the attacks less discernible, however they were still perceived as noise in the recording. It can be theorized that increasing \emph{$\lambda$} further may eventually make these attacks less noticeable, however our studies have shown that the number of iterations required to generate an attack grows exponentially with larger \emph{$\lambda$} and we were unable to reach the point where such attacks can be considered imperceptible. We have also found that despite our predictions, applying PAM to the attack generation process has led to a substantial increase in effectiveness in over the air scenarios.

\section{Acknowledgements}\label{acknowledgements}

I would like to thank my supervisor Vasiliy Dolmatov for his guidance and support, which greatly helped my research.

\clearpage
\printbibliography

\end{document}